\documentclass{article}
\usepackage[T1]{fontenc} 
\usepackage[utf8]{inputenc} 
\usepackage{ismir,amsmath,cite,url}
\usepackage{graphicx}
\usepackage{color}

\usepackage{lineno}
\usepackage[labelsep=period]{caption}

\usepackage{multirow,caption,subcaption,changepage,arydshln}

\title{Self-refining of Pseudo Labels\\for Music Source Separation with Noisy Labeled Data}

\multauthor
{*Junghyun Koo$^1$\thanks{*Equal contribution} \hspace{1cm} *Yunkee Chae$^2$ \hspace{1cm} Chang-Bin Jeon$^1$ \hspace{1cm} Kyogu Lee$^{1,2,3}$} {
$^1$Department of Intelligence and Information, 
$^2$Interdisciplinary Program in Artificial Intelligence, \\
$^3$Artificial Intelligence Institute, Seoul National University \\
{\tt\small \{dg22302, yunkimo95, vinyne, kglee\}@snu.ac.kr}
}

\def\authorname{J. Koo, Y. Chae, C.-B. Jeon, and K. Lee}

\usepackage[bookmarks=false,pdfauthor={\authorname},pdfsubject={\papersubject},hidelinks]{hyperref}

\begin{document}

\maketitle

\begin{abstract}
Music source separation (MSS) faces challenges due to the limited availability of correctly-labeled individual instrument tracks. 
With the push to acquire larger datasets to improve MSS performance, the inevitability of encountering mislabeled individual instrument tracks becomes a significant challenge to address.
This paper introduces an automated technique for refining the labels in a partially mislabeled dataset. Our proposed self-refining technique, employed with a noisy-labeled dataset, results in only a 1\% accuracy degradation in multi-label instrument recognition compared to a classifier trained on a clean-labeled dataset.
The study demonstrates the importance of refining noisy-labeled data in MSS model training and shows that utilizing the refined dataset leads to comparable results derived from a clean-labeled dataset.
Notably, upon only access to a noisy dataset, MSS models trained on a self-refined dataset even outperform those trained on a dataset refined with a classifier trained on clean labels.
\end{abstract}

\begin{figure}[t]
  \centering
  \includegraphics[width=\linewidth]{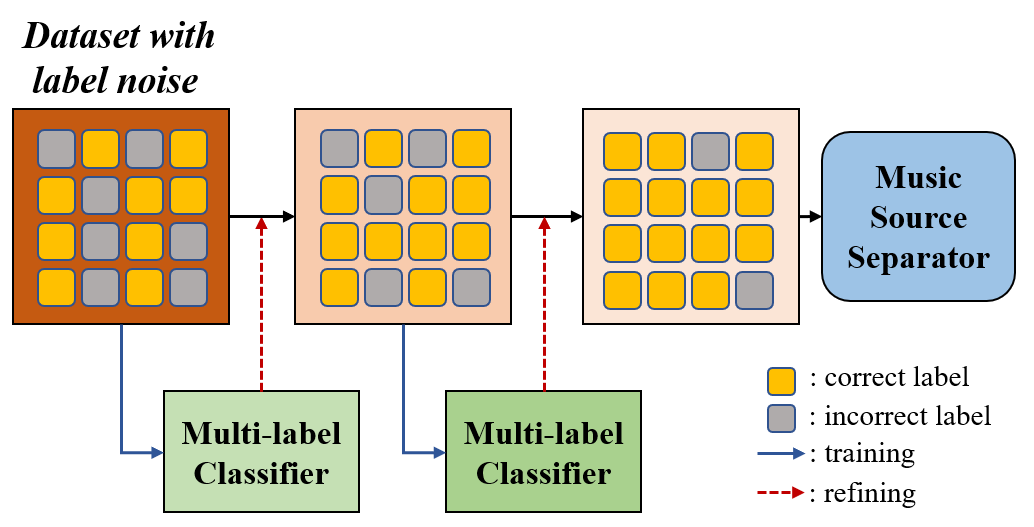}
  \caption{Overview of self-refining procedure on a noisy-labeled dataset for music source separation.}
  \label{fig:overall_framework}
\vspace{-10pt}
\end{figure}

\section{Introduction}

Music source separation (MSS) is a critical task in the field of music information retrieval (MIR), with applications ranging from remixing \cite{woodruff2006remixing, pons2016remixing, koo2023music} to transcription \cite{plumbley2002automatic, lin2021unified, cheuk2023jointist} and music education \cite{dittmar2012music, cano2014pitch}. To train high-performing MSS models, it is essential to have clean single-stem music recordings for guidance, which serve as the ground truth for model training. However, obtaining clean, large-scale datasets of single instrument tracks remains a challenging task.

With the increasing availability of music data on the internet, platforms such as YouTube provide a vast pool of potential single-instrument tracks. Although these sources offer an opportunity for performance gains through larger training datasets, collecting single instrument tracks from such platforms inevitably leads to encountering tracks with incorrect labels. For example, a query aimed at obtaining drum recordings might yield results that contain other types of instruments or noise, causing discrepancies between the expected and actual content of the collected recordings.

\textbf{Label noise} in datasets can arise from various factors, such as bleeding between instrument tracks, mislabeling due to human error, or the ambiguous timbre of instruments that resemble other instrument categories \cite{hoopen1994issues}. These factors make it challenging to assign a single definitive instrument label to a given recording. Such label noise is detrimental to the performance of MSS models, and there is a pressing need for an approach that can effectively train MSS models using partially corrupted datasets.

In response to this challenge, we propose an automated approach for refining mislabeled instrument tracks in a partially noisy-labeled dataset. Our \textit{self-refining} technique, which leverages noisy-labeled data, results in only a 1\% accuracy degradation for multi-label instrument recognition compared to a classifier trained with a clean-labeled dataset. The study highlights the importance of refining noisy-labeled data for training MSS models and demonstrates that utilizing the refined dataset for MSS yields results comparable to those obtained using a clean-labeled dataset. Notably, when only a noisy dataset is available, MSS models trained on self-refined datasets even outperform those trained on datasets refined with a classifier trained on clean labels. This paper presents a comprehensive analysis of our proposed method and its impact on the performance of MSS models.

\begin{figure*}[t]
  \centering
  \includegraphics[width=\linewidth]{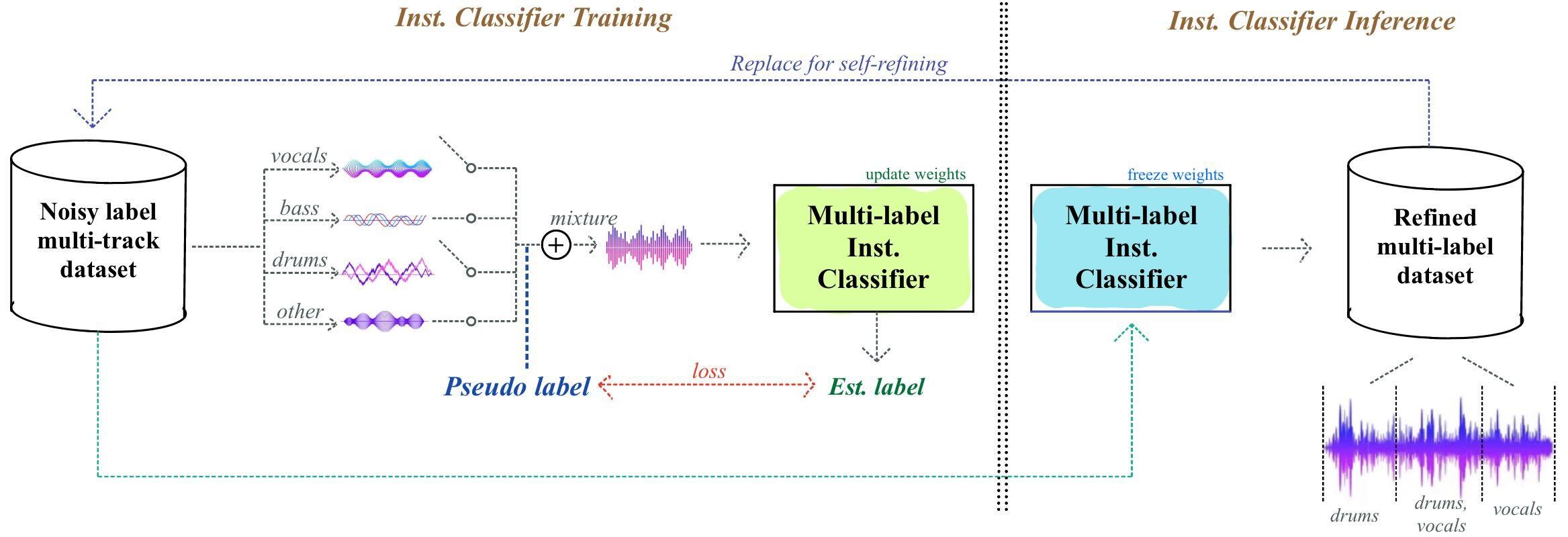}
  \caption{Overall training procedure of the Instrument Classifier $\Psi$. The classifier is trained to perform instrument recognition with mixtures that are synthesized by randomly selecting each stem from the noisy labeled dataset. After this training procedure, we refine the original noisy dataset and then use this new dataset to train the final $\Psi$.
}
  \label{fig:model_pipeline_cla}
\vspace{-5pt}
\end{figure*}

\section{Related Works}
\label{sec:related}
\textbf{Self-training} of machine learning models has been studied in various literatures, where a teacher model is first trained with clean labeled data and is used as a label predictor of unlabeled data, then a student model is trained with clean and pseudo-labeled data \cite{scudder1965probability, yarowsky1995unsupervised}. Recently, Xie et al. proposed a noisy student method for self-training \cite{xie2020self}, which uses an iterative training of teacher-student models and noise injection methods for training student models. Thanks to their usefulness, these self-training methods have been used in diverse MIR tasks, such as singing voice detection \cite{schluter2016learning, meseguer2019dali} and vocal melody extraction \cite{keum2020semi}.

\textbf{Instrument recognition} or classification has been researched in various literatures, both in single-instrument \cite{benetos2006musical, eronen2000musical, lostanlen2016deep, essid2006hierarchical} or multi-source settings \cite{han2016deep, hung2018frame, gururani2018instrument, gururani2019attention, hung2019multitask, flores2021leveraging, reghunath2022transformer, kim2023show}. Although such research has been focused on single or predominant-label prediction, Zhong, et al. \cite{zhong2023attention} recently proposed the hierarchical approach for multi-label music instrument classification.

Our self-refining method for training of instrument classifier shares similar attributes with noisy student training \cite{xie2020self} and the previous multi-label instrument classification \cite{zhong2023attention} but differs from some perspectives. 
\textit{\romannumeral1)} We train all our models only with partially noisy-labeled data, without access to clean-labeled data.
\textit{\romannumeral2)} We train the classifiers for direct prediction of labels used in standard music source separation, e.g., vocals, bass, drums, and others, instead of the hierarchical approach. 
\textit{\romannumeral3)} 
We train multi-label classifiers with mixtures of randomly selected instruments, which are based on the characteristic of musical audio.
If there exist two different instruments in one audio signal, that can be classified into two instruments. This random mixing of different instrumental tracks has been used in music source separation as well \cite{uhlich2017improving}. Note that the mixup method \cite{zhang2017mixup}, which is also a mixing method of two different images, also shares a similar attribute with our method but is used for regularization of training single-label classifiers, not like our multi-label classifiers.

\begin{figure*}[t]
  \centering
  \includegraphics[width=0.75\linewidth]{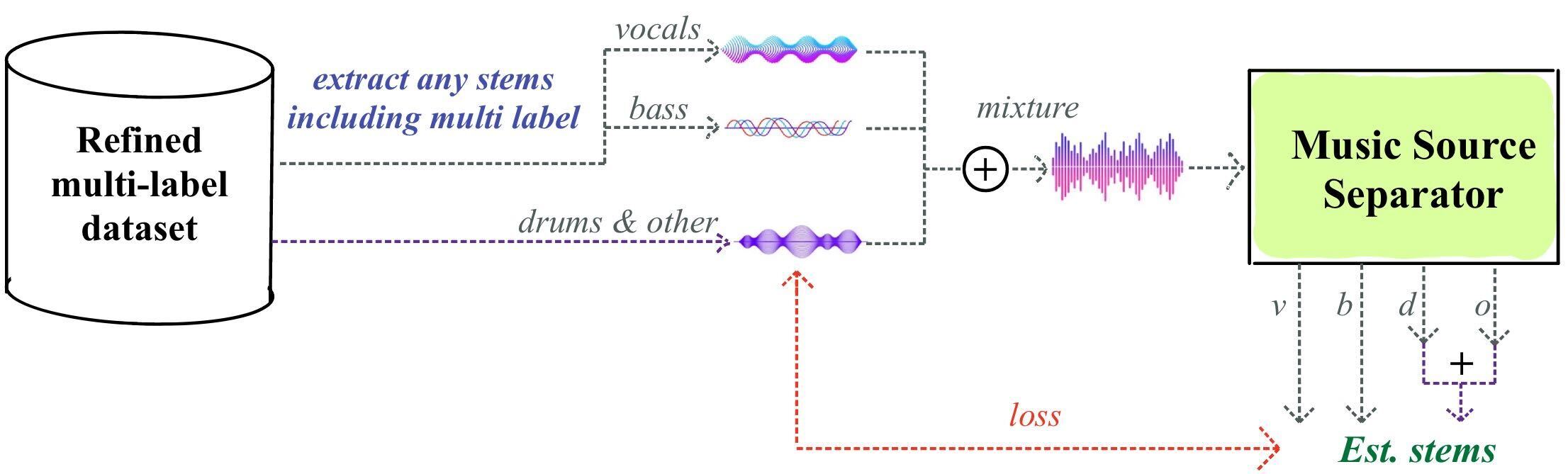}
  \caption{Music source separation training. Similar to the training procedure of the instrument classifier, we randomly mix each stem from the refined dataset to synthesize a mixture and use it as a network input. When a multi-labeled segment is selected for synthesis, the corresponding estimated stems are summed for loss computation.
}
  \label{fig:model_pipeline_mss}
\vspace{-10pt}
\end{figure*}

\section{Methodology}
\label{sec:methodology}
Given a real-world scenario where the available multi-track dataset for MSS is partially incorrect with its instrument labels, a possible naive approach is first to rectify mislabeled tracks and then train an MSS model using stems with the revised labels.
In this section, we introduce an effective training technique that first performs instrument recognition by only utilizing data with noisy labels and then leverages the refined dataset inferred with the trained multi-label instrument classifier to train the MSS model.
With this two-stage approach, we explore the impact of the refined noisy datasets on the performance of MSS models.

\subsection{Multi-label Instrument Recognition}
\label{subsec:method_cla}

Figure \ref{fig:model_pipeline_cla} summarizes the proposed training procedure of the Instrument Classifier $\Psi$.
Similar yet different from self-training, our approach learns directly from noisy labeled data and re-labels the training data to train the final $\Psi$ using this refined dataset.
We call this training procedure \textit{self-refining}, and this is possible by \textit{random mixing}, a method to synthesize a mixture of multiple instruments with pseudo labels.
The \textit{random mixing} technique takes advantage of the acoustic music domain in that mixing sources of different instrument tracks still leads to natural output mixture, whereas naively combining different images in the image domain is likely to produce unrealistic results.
We further discuss about the benefits and the detailed process of \textit{random mixing} at \ref{subsubsec:rand_mix}.

The network architecture of $\Psi$ is that of the ConvNeXt model \cite{liu2022convnet}, where it has shown great performance on a multi-instrument retrieval in \cite{kim2023show}. 
The input of the network is a stereo-channeled magnitude linear spectrogram. Followed by a sigmoid layer, the model outputs four labels indicating the presence of each stem. 
The objective function for instrument recognition $\mathcal{L}_{\Psi}$ is a mean absolute loss between the estimated and synthesized pseudo labels.
Preliminary experiments showed no significant difference in performance when employing mean absolute loss as compared to binary cross-entropy loss. This is likely due to the random mixing sampling that ensures similar occurrences of positive and negative labels of each instrument class during the training procedure.

\subsubsection{Random Mixing}
\label{subsubsec:rand_mix}
Randomly mixing stems with label noise not only creates various combinations of multi-labeled mixtures for training the instrument classifier but also brings the chance to generate a correct pseudo label from mislabeled stems. For instance, if we randomly select one correctly labeled drum track and a track that contains both sources of drums and vocals but is mislabeled as vocals, the mixing process synthesizes a correctly labeled mixture.
Thanks to these fortunate chances, the random mixing technique assists the instrument classifier's accuracy in refining the label noise dataset by utilizing mislabeled stems.

To synthesize a random mixture and its pseudo label, each stem is first selected with a chance rate from the noisy dataset. 
The audio effects manipulation is then applied to each chosen track by simulating the music mixing process for data augmentation \cite{koo2023music}. The order of applying audio effects with random parameters is 1. \textit{dynamic range compression}, 2. \textit{algorithmic reverberation}, 3. \textit{stereo imaging}, and 4. \textit{loudness manipulation}.
Labels corresponding to randomly selected stems are used as a multi-label objective for the instrument classifier.

\subsection{Music Source Separation}
\label{subsec:method_mss}
In this section, we describe the training procedure of MSS model employing a multi-labeled refined dataset curated by the classifier trained in Section \ref{subsec:method_cla}.
The majority of MSS research has focused on estimating each of the four instrument groups (\textit{vocals}, \textit{bass}, \textit{drums}, and \textit{other}) \cite{openumx, mdxnet, htdemucs, bsrnn}.
However, our refined dataset contains sources labeled with multiple stems, which are unsuitable for use as distinct target instruments. 
To utilize multi-labeled sources, we propose an appropriate MSS training framework tailored to our refined dataset. 

First, we determine whether to include the multi-stem source for each input mixture sample by considering the probability $p$.
If we decide not to include the multi-labeled source, we can train the MSS model in a conventional manner, computing the losses for each stem. 
Otherwise, we select a multi-labeled source from the refined dataset.
Subsequently, we choose the remaining stems that do not correspond to the selected multi-labeled source from a pool of single-labeled sources and combine them to simulate a mixture.
For example, when selecting a multi-labeled source \textit{bass+drums}, we opt for single sources labeled as \textit{vocals} and \textit{others} to synthesize the mixture.
After conducting inference with the MSS model, we add the estimated stems corresponding to the multi-stem source of the input mixture and assess the loss between them.
Figure \ref{fig:model_pipeline_mss} illustrates our training procedure when a multi-labeled source is selected.
We compute the losses for each stem, treating the multi-labeled source as an individual stem, and subsequently sum these losses to derive the final loss value.

\begin{table*}[]
\begin{tabular}{ccccccc}
\hline
\multirow{2}{*}{\textbf{Label Type}} & \multirow{2}{*}{\textbf{Training Data}} & \multicolumn{5}{c}{\textbf{\begin{tabular}[c]{@{}c@{}}Accuracy / F1 Score\\ Precision / Recall\end{tabular}}}                                                                                                                                                                                                                                                                                                   \\ \cline{3-7} 
                                     &                                         & \textit{vocals}                                                               & \textit{bass}                                                                 & \textit{drums}                                                                & \textit{other}                                                                  & \textit{avg}                                                                  \\ \hline
\multirow{5}{*}{Single-Label}        & \textit{clean}                          & \begin{tabular}[c]{@{}c@{}}97.8\% / 0.947\\ 0.91 / 0.98\end{tabular}          & \begin{tabular}[c]{@{}c@{}}94.4\% / 0.891\\ 0.84 / 0.94\end{tabular}          & \begin{tabular}[c]{@{}c@{}}95.1\% / 0.914\\ 0.85 / 0.98\end{tabular}          & \begin{tabular}[c]{@{}c@{}}93.2\% / 0.880\\ 0.90 / 0.85\end{tabular}            & \begin{tabular}[c]{@{}c@{}}95.1\% / 0.906\\ 0.87 / 0.93\end{tabular}          \\ \cdashline{2-7}
                                     & \textit{noisy}                          & \begin{tabular}[c]{@{}c@{}}93.6\% / 0.860\\ 0.76 / 0.97\end{tabular}          & \begin{tabular}[c]{@{}c@{}}\textbf{90.0\% / 0.821}\\ \textbf{0.73} / 0.93\end{tabular} & \textbf{\begin{tabular}[c]{@{}c@{}}93.7\% / 0.893\\ 0.81 / 0.98\end{tabular}} & \begin{tabular}[c]{@{}c@{}}\textbf{92.6\% / 0.865}\\ \textbf{0.92} / 0.81\end{tabular} & \begin{tabular}[c]{@{}c@{}}92.5\% / 0.860\\ \textbf{0.80} / 0.92\end{tabular}          \\
                                     & \textit{refined}                        & \textbf{\begin{tabular}[c]{@{}c@{}}96.1\% / 0.911\\ 0.84 / 0.98\end{tabular}} & \begin{tabular}[c]{@{}c@{}}89.6\% / 0.818\\ 0.71 / \textbf{0.96}\end{tabular}          & \begin{tabular}[c]{@{}c@{}}93.1\% / 0.884\\ 0.79 / \textbf{0.98}\end{tabular}          & \begin{tabular}[c]{@{}c@{}}92.3\% / 0.862\\ 0.90 / \textbf{0.82}\end{tabular}            & \textbf{\begin{tabular}[c]{@{}c@{}}92.8\% / 0.866\\ 0.80 / 0.93\end{tabular}} \\ \hline
\multirow{5}{*}{Multi-Label}         & \textit{clean}                          & \begin{tabular}[c]{@{}c@{}}92.4\% / 0.929\\ 0.92 / 0.93\end{tabular}          & \begin{tabular}[c]{@{}c@{}}89.6\% / 0.905\\ 0.89 / 0.92\end{tabular}          & \begin{tabular}[c]{@{}c@{}}90.5\% / 0.913\\ 0.87 / 0.95\end{tabular}          & \begin{tabular}[c]{@{}c@{}}88.1\% / 0.878\\ 0.90 / 0.85\end{tabular}            & \begin{tabular}[c]{@{}c@{}}90.2\% / 0.907\\ 0.90 / 0.91\end{tabular}          \\ \cdashline{2-7}
                                     & \textit{noisy}                          & \begin{tabular}[c]{@{}c@{}}87.9\% / 0.895\\ 0.83 / 0.96\end{tabular}          & \begin{tabular}[c]{@{}c@{}}87.5\% / 0.888\\ \textbf{0.86} / 0.93\end{tabular}          & \begin{tabular}[c]{@{}c@{}}87.7\% / 0.891\\ 0.82 / \textbf{0.96}\end{tabular}          & \begin{tabular}[c]{@{}c@{}}87.3\% / 0.872\\ \textbf{0.88 / 0.87}\end{tabular}            & \begin{tabular}[c]{@{}c@{}}87.6\% / 0.887\\ 0.85 / 0.93\end{tabular}          \\
                                     & \textit{refined}                        & \textbf{\begin{tabular}[c]{@{}c@{}}91.9\% / 0.928\\ 0.88 / 0.97\end{tabular}} & \begin{tabular}[c]{@{}c@{}}\textbf{87.8\% / 0.894}\\ 0.84 / \textbf{0.95}\end{tabular} & \textbf{\begin{tabular}[c]{@{}c@{}}89.6\% / 0.906\\ 0.85 / 0.96\end{tabular}} & \textbf{\begin{tabular}[c]{@{}c@{}}87.4\% / 0.874\\ 0.88 / 0.87\end{tabular}}   & \textbf{\begin{tabular}[c]{@{}c@{}}89.2\% / 0.901\\ 0.86 / 0.94\end{tabular}} \\ \hline
\end{tabular}
\vspace{-5pt}
\caption{
    Single and multi-label instrument recognition performance of instrument classifiers trained with different datasets.
    The training data of \textit{clean}, \textit{noisy}, and \textit{refined} each represents the training subset of MUSDB18, MDX2023, and MDX2023 refined with the instrument classifier trained with MDX2023 $\Psi_{\textit{noisy}}$, respectively.
}
\vspace{-10pt}
\label{table:inst_cla_obj_results}
\end{table*}

\section{Experiments}
\label{sec:experiments}

\subsection{Dataset}
We use the label noise dataset provided by the Music Demixing Challenge 2023 (MDX2023) \cite{aicrowd}, which consists of 203 songs, licensed by \textit{Moises.AI}\footnote{https://moises.ai/}.
Similar to MUSDB18 \cite{musdb18-hq}, the provided dataset contains mixtures of music recordings segregated into four different instrumental stems: \textit{vocals}, \textit{bass}, \textit{drums}, and \textit{other}.
Each stem and its corresponding label are intentionally altered to produce a corrupted dataset to simulate mislabeling such as bleeding or human mistakes.
That is, for instance, \texttt{drums.wav} may contain drum sounds and singing voices simultaneously, which is likely to be caused by bleeding.
For another example, a kick-drum sound might be mislabeled as \texttt{bass.wav} when the pitch of the kick drum is melodic enough to trick a human labeler. 
Due to the nature of the MDX2023 challenge, the dataset does not contain the actual ground truth labels. Hence, we use all 203 songs of the MDX2023 dataset only as training data.

To validate our system trained with noisy labeled data, we employed the MUSDB18 \cite{musdb18-hq} as the clean dataset for comparison and evaluation.
MUSDB18 comprises 150 songs, with 100 songs for the training and 50 songs for the test set. We adopt the test subset for evaluating all systems, while the training subset is used to train the upper bound system for observation.

\textbf{Data preprocessing.} To prevent models from mislabels caused by silence, we remove all silent sections throughout both datasets. The preprocessing procedure for silence removal is as follows:
\begin{enumerate}
    \item For each song, detect silent areas that are below 30 dB relative to the maximum peak amplitude.
    \item Remove all detected areas then merge them into one single long audio track.
    \item Repeat 1. (with the threshold of 60 dB) and 2. based on the merged audio track, in case of stems that are almost silent.
\end{enumerate}
After trimming silent regions, the total durations for each stem in the respective order of \textit{vocals}, \textit{bass}, \textit{drums}, and \textit{other} are [7.2, 7.8, 9.2, 10.3] hours for the MDX2023 dataset, and [2.2, 2.7, 2.9, 3.3] hours for the test subset of MUSDB18. 
Note that for evaluating MSS performance, we instead follow the original convention of processing entire songs from the test subset without any silence removal. 
We use the original audio specifications of both datasets where all audio tracks are stereo-channeled and have a sampling rate of 44.1 KHz.

\subsection{Experimental Setups}
\label{exp_setup}
For multi-label instrument recognition, the network architecture of $\Psi$ is ConvNeXt's tiny version \cite{liu2022convnet}, which consists of 27.8M parameters. We feed the network with stereo-channeled mixtures of instruments that are of 2.97 seconds, which are transformed into a time-frequency domain linear magnitude spectrogram with an FFT size of 2048 and a hop size of 512. We train all $\Psi$ for 100 epochs.
During inference, $\Psi$ performs classification by processing the entire input audio in windows of a size equivalent to the network input size, with a hop size of one-fourth of this window size. The output labels from these windows are then averaged to yield the final decision, based on a threshold value of 0.9.
We utilized this inference procedure to refine the noisy dataset, which was then used to train our MSS models. Our final version of the instrument classifier trained on the refined dataset $\Psi_{\textit{refined}}$ only uses stems inferred as a single-labeled for better performance based on our preliminary experiments.

We employed two MSS models, Hybrid Demucs (Demucs v3) \cite{hdemucs} and CrossNet-Open-Unmix (X-UMX) \cite{xumx}, to evaluate their performance when trained on the processed datasets. 
Multi-labeled sources were selected with a probability of 0.4, and input loudness normalization (-14 LUFS) was applied for both training and inference in accordance with \cite{jeon2022towards}. \texttt{pyloudnorm} \cite{steinmetz2021pyloudnorm} was used for loudness calculation \cite{itu2011itu}.

For Demucs, the input duration was set to 3 seconds, and optimization was performed using Adam optimizer \cite{kingma2014adam} and L1 loss on the time domain.
The model was trained for 21,000 iterations with a batch size of 160.

For X-UMX, the input duration was set to 6 seconds, and optimization was performed using AdamW optimizer \cite{loshchilov2017decoupled} and mean squared error loss on the time-frequency domain. For the sake of simplicity, we omit the multi-domain and combination loss proposed in \cite{xumx}. The model was trained for 56,400 iterations with a batch size of 32. For the \textit{+ finetune w/ multi-labeled} model in Table \ref{table:mss_ablation}, we first train the model with only single-labeled data for 20,680 iterations, then finetune it with multi-labeled data for another 35,720 iterations.

\section{Results}
\label{sec:results}

\subsection{Instrument Recognition}
\label{subsec:result_cla}
Table \ref{table:inst_cla_obj_results} presents the instrument recognition performance of the multi-instrument classifier on single-labeled and multi-labeled data.
As ground-truth labels are not available for the MDX2023 dataset, we validate the classification performance according to single and multi-labeled data with the MUSDB18 test set for evaluation.
For the multi-label evaluation, we synthesized 3,941 mixtures from the test set with the random mixing technique described in \ref{subsubsec:rand_mix}. 
We observe the performance of $\Psi$ trained with MUSDB18 (\textit{clean}), MDX2023 (\textit{noisy}), and MDX2023 once refined with $\Psi_{\textit{noisy}}$ (\textit{refined}).
The evaluation metrics used are accuracy, F1 score, precision, and recall for each instrument class and the overall averaged result.

For single-labeled data, the classifier achieves the highest average performance on the \textit{clean} dataset, with an accuracy of 95.1\% and an F1 score of 0.906. 
As \textit{clean} dataset does not contain any noisy labels, the obtained results can be considered an upper bound for the performances of the classifiers.
The $\Psi$ trained on \textit{refined} dataset results in slightly lower performance, with an accuracy of 92.8\% and F1 score of 0.866, while the \textit{noisy} dataset shows an accuracy of 92.5\% and F1 score of 0.860.
Although the accuracy, F1 score, and precision are higher for the \textit{noisy} dataset in the \textit{bass, drums}, and \textit{other} stems, the performance metrics for \textit{vocals} and recall values across all stems exhibit superior results when trained with the \textit{refined} dataset.

For instrument recognition on multi-labeled data, $\Psi$ trained on \textit{clean} dataset yields an average accuracy of 90.2\% and F1 score of 0.907. 
The \textit{noisy} dataset results in an accuracy of 87.6\% and an F1 score of 0.887.
The \textit{refined} dataset achieves superior performance, with an accuracy of 89.2\% and an F1 score of 0.901, which is comparable to the results obtained from the \textit{clean} dataset.
Contrary to the evaluation with single-labeled data, the \textit{refined} dataset generally demonstrates superior performance across all metrics in comparison to the \textit{noisy} dataset. Notably, the recall values are observed to be even higher than those of the \textit{clean} dataset.
An in-depth analysis of the multi-instrument classifier results, alongside the performance outcomes of the MSS models, is discussed in Section \ref{subsec:result_mss}.

\begin{table}[]
\centering
\setlength\tabcolsep{2.5pt}
\begin{tabular}{clccccc}
\hline
\multirow{2}{*}{\textbf{Network}} & \multicolumn{1}{c}{\multirow{2}{*}{\textbf{\begin{tabular}[c]{@{}c@{}}Training\\ Data\end{tabular}}}} & \multicolumn{5}{c}{\textbf{SDR {[}dB{]}}}                                                                                                                                  \\ \cline{3-7} 
                                  & \multicolumn{1}{c}{}                                                                                  & \textit{vocals}                   & \textit{bass}                     & \textit{drums}                    & \textit{other}                    & \textit{avg}                      \\ \hline
\multirow{5}{*}{\begin{tabular}[c]{@{}c@{}} \textit{Demucs}\\ \cite{hdemucs}\end{tabular}}  & \textit{clean}                                                                                        & \multicolumn{1}{c}{5.92}          & \multicolumn{1}{c}{6.16}          & \multicolumn{1}{c}{5.58}          & \multicolumn{1}{c}{4.43}          & \multicolumn{1}{c}{5.52}          \\ \cdashline{2-7}
                                  & \textit{noisy}                                                                                        & \multicolumn{1}{c}{3.37}          & \multicolumn{1}{c}{1.92}          & \multicolumn{1}{c}{0.70}          & \multicolumn{1}{c}{0.86}          & \multicolumn{1}{c}{1.71}          \\
                                  & \textit{\; w/ $\Psi_{\text{clean}}$}                                                                   & \multicolumn{1}{c}{5.31} & \multicolumn{1}{c}{\textbf{5.12}}          & \multicolumn{1}{c}{1.32}          & \multicolumn{1}{c}{2.16}          & \multicolumn{1}{c}{3.48}          \\
                                  & \textit{\; w/ $\Psi_{\text{noisy}}$}                                                                   & \multicolumn{1}{c}{4.15}          & \multicolumn{1}{c}{4.58}          & \multicolumn{1}{c}{1.62}          & \multicolumn{1}{c}{2.85} & \multicolumn{1}{c}{3.30}          \\
                                  & \textit{\; w/ $\Psi_{\text{refined}}$}                                                                 & \multicolumn{1}{c}{\textbf{5.36}}          & \multicolumn{1}{l}{5.04} & \multicolumn{1}{c}{\textbf{3.09}} & \multicolumn{1}{c}{\textbf{3.13}}          & \multicolumn{1}{c}{\textbf{4.16}} \\ \hline
\multirow{5}{*}{\begin{tabular}[c]{@{}c@{}}\textit{X-UMX}\\ \cite{xumx}\end{tabular}}   & \textit{clean}                                                                                        & 5.76                              & 4.44                              & 5.47                              & 3.65                              & 4.83                              \\ \cdashline{2-7}
                                  & \textit{noisy}                                                                                        & 3.39                              & 1.78                              & 1.52                              & 0.96                              & 1.91                              \\
                                  & \textit{\; w/ $\Psi_{\text{clean}}$}                                                                   & 4.50                              & 3.22                              & 3.66                              & 2.73                              & 3.53                              \\
                                  & \textit{\; w/ $\Psi_{\text{noisy}}$}                                                                   & 4.72                              & \textbf{4.11}                     & 3.22                              & 2.89                              & 3.74                              \\
                                  & \textit{\; w/ $\Psi_{\text{refined}}$}                                                                 & \textbf{4.99}                     & 3.93                              & \textbf{5.00}                     & \textbf{3.18}                     & \textbf{4.28}                     \\ \hline
\end{tabular}
\caption{Source separation performance of Demucs v3 \cite{hdemucs} and CrossNet-Open-Unmix \cite{xumx} trained on different training datasets. Sub-items below \textit{noisy} dataset indicate data refined with the respective instrument classifiers, denoted as $\Psi_{\bullet}$.
}
\vspace{-10pt}
\label{table:mss_obj_result}
\end{table}

\begin{figure*}[t]
    \centering
    \includegraphics[width=\textwidth]{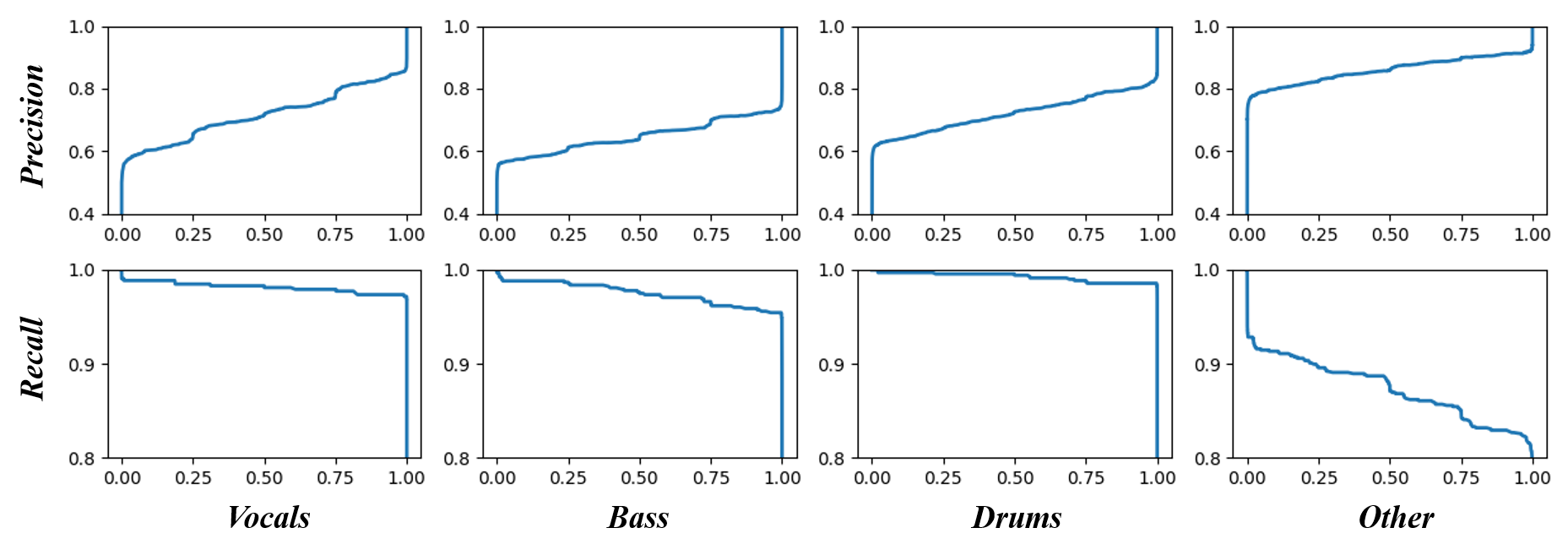}
    \setlength{\abovecaptionskip}{-10pt}
    \caption{Precision and recall curves of the proposed classifier across different thresholds (x-axis) on each instrument.
    The curves are generated using the MUSDB18 test set (\textit{clean}).}
    \label{fig:prcurves}
\vspace{-13pt}
\end{figure*}

\subsection{Source Separation}
\label{subsec:result_mss}
The results of MSS models trained on different training datasets are presented in Table \ref{table:mss_obj_result}. 
In our evaluation, we used Signal-to-Distortion Ratio (SDR) \cite{vincent2006performance}, which is calculated using the \texttt{museval} toolkit \cite{museval}. For all MSS experiments, we report the SDR median of frames and the median of tracks. 
The Demucs and X-UMX models are trained on \textit{clean}, \textit{noisy}, and data processed with multi-instrument classifiers, denoted by $\Psi_{\bullet}$.
In this context, $\Psi_{\bullet}$ represents the classifier trained on each respective dataset, as described in Section \ref{subsec:result_cla}.

The baseline for this experiment is established using MSS models trained on the \textit{noisy} dataset.
It is noteworthy that all the results presented in the table exceed the baseline performance.
For the dataset processed with the multi-instrument classifier $\Psi_{\textit{refined}}$, average SDR improvements of 2.45 and 2.31 are observed for Demucs and X-UMX models, respectively, in comparison to the \textit{noisy} dataset.
Specifically, in $\Psi_{\textit{refined}}$ case, both Demucs and X-UMX models demonstrate substantial improvements in SDR values across all stems compared to those of $\Psi_{\textit{noisy}}$, with the exception of \textit{bass} in the X-UMX model.

\begin{table}[]
\centering
\setlength\tabcolsep{2.5pt}
\begin{tabular}{lccccc}
\hline
\multicolumn{1}{c}{\multirow{2}{*}{\textbf{Method}}} & \multicolumn{5}{c}{\textbf{SDR {[}dB{]}}}                                 \\ \cline{2-6} 
\multicolumn{1}{c}{}                                 & \textit{vocals} & \textit{bass} & \textit{drums} & \textit{other} & \textit{avg} \\ \hline
\textit{proposed}                                    & 4.99            & 3.93          & 5.00           & 3.18           & 4.28         \\ \cdashline{1-6}
\textit{threshold = 0.5}                             & 5.06            & 4.13          & 4.77           & 3.06           & 4.25         \\
\textit{adaptive thresholds}  & 4.70 & 3.72 & 3.70 & 2.62 & 3.68 \\ \cdashline{1-6}
\fontsize{9}{9}\selectfont\textit{train only w/ single-labeled}        & 4.90            & 3.73          & 4.54           & 3.18           & 4.09         \\
\fontsize{9}{9}\selectfont\textit{+ finetune w/ multi-labeled}            & 4.33            & 4.33          & 4.19           & 3.14           & 4.00         \\ \cdashline{1-6}
\fontsize{9}{9}\selectfont\textit{self-refining $\times5$}        & 4.65            & 3.87          & 5.07           & 2.89           & 4.12         \\\hline
\end{tabular}
\caption{Ablation studies on MSS performances with CrossNet-Open-Unmix.}
\vspace{-10pt}
\label{table:mss_ablation}
\end{table}

\subsubsection{Analysis in relation to instrument recognition}
\label{subsubsec:analyze_ir}
In Table \ref{table:mss_obj_result}, it is noteworthy that the performance of $\Psi_{\textit{refined}}$ exceeds the performance of $\Psi_{\textit{clean}}$, even though $\Psi_{\textit{clean}}$ is trained with a noise-free labeled dataset.
This implies the classification performance of $\Psi_{\textit{clean}}$ is inferior to the classification performance of $\Psi_{\textit{refined}}$.
This discrepancy could be attributed to differences in the data distribution between the MUSDB18 and MDX2023 datasets.
Moreover, the number of training samples varies, with 100 samples in the MUSDB18 dataset and 203 samples in the MDX2023 dataset.
When refining a partially noisy dataset, employing the same partially noisy dataset can yield advantageous outcomes than using the smaller clean dataset. This observation might be aligned with the findings in \cite{xie2020self}, which report an improvement in performance when a larger quantity of unlabeled data is present.

An additional factor to consider is the distinctive nature of the MSS model training framework in our approach. 
MSS models utilize the output of the classifier as input.
The performance of the MSS model can be affected differently depending on the type of error in the classifier's output.
For example, assume that the MSS model receives a sample misclassified as a vocal stem when no vocals are actually present (i.e. a false-positive sample for vocals).
In this case, the MSS model simply needs to predict silence for the vocals stem and produce it as output, resulting in no significant confusion.
Conversely, consider a scenario in which the MSS model receives a sample misclassified as a non-vocal stem (e.g. drums + bass), despite the presence of vocals, resulting in a false-negative sample for vocals.
In such a case, the model will attempt to allocate the vocals present in the input data to the drum and bass stems. 
Furthermore, our model differs from traditional MSS training methods as it also accepts multi-stem data as input. 
In this context, the vocals are present as the correct answer for multiple mislabeled non-vocal stems, which confuses the model. 
This not only negatively affects the performance of the mislabeled stems but also the vocal stem itself.

As a consequence of the unique characteristics of our training process, false-negative samples have a more significant impact on MSS compared to false-positive samples, highlighting the increased significance of the recall metric.
Considering this perspective, the results presented in Table \ref{table:inst_cla_obj_results} imply the possibility of the sub-optimal performance of MSS trained on outputs of $\Psi_{\textit{clean}}$, where the recall values are lower for all stems compared to $\Psi_{\textit{refined}}$.

\subsubsection{Ablation studies}
\label{subsubsection:ablation}
As shown in Table \ref{table:mss_ablation}, we evaluate the performance of X-UMX under various conditions to better understand the significance of distinct aspects of our proposed method.

\noindent\textbf{Threshold.} We conduct experiments to examine the impact of threshold determination for the classifier during the training of MSS models using a classified dataset. The evaluation is performed on the MUSDB18 test set. We observe that reducing the threshold to 0.5 only exhibits an SDR of 0.03 degradation compared to the original threshold value of 0.9.
This outcome can be attributed to the fact that only 8\% of $\Psi_{\textit{refined}}$ outputs fall within the range of [0.1, 0.9] upon inference on the MUSDB18 test set.
In Figure \ref{fig:prcurves}, we present the precision and recall curves for each threshold on individual instruments.
It is evident from the curves that the variations within that range for both precision and recall are not substantial.
Consequently, the choice between thresholds of 0.9 or 0.5 does not yield any noticeable disparity.
Furthermore, we conduct an experiment involving adaptive thresholds for each instrument, where the threshold for each instrument was set to maximize the F1 score of the classification performance. 
However, we observe a significant degradation in performance across all instruments when employing adaptive thresholds.
Maximizing the F1 score necessitates a trade-off between recall and precision, often leading to a decline in recall to enhance precision.
Consequently, the performance of the MSS model experience degradation, aligning with the discussion presented in Section \ref{subsubsec:analyze_ir}.

\noindent\textbf{Training with multi-labeled data.} When training solely with the data estimated as single-labeled, the performance is not as good as that of the proposed method.
Incorporating both single- and multi-labeled data for fine-tuning after the initial training on single-labeled data leads to a slightly diminished performance, despite utilizing both types of labeled data during the training process.

\noindent\textbf{Iterative self-refining.}
Finally, we examine the influence of the iterative self-refining technique on MSS performance.
The results indicate that an MSS model trained with a noisy-labeled dataset refined five times through our method does not yield superior performance compared to the proposed model, trained on a dataset refined twice, and the performance difference is insignificant.
This observation suggests that excessive refinement iterations do not necessarily lead to improved performance and that refining the dataset twice may be sufficient for optimal results.

\section{Conclusion}
\label{section:conclusion}
In conclusion, this paper presented a self-refining approach to address the challenges of noisy-labeled data in training music source separation (MSS) models. Our proposed method refines mislabeled instrument tracks in partially noisy-labeled datasets, resulting in only a 1\% accuracy degradation for multi-label instrument recognition compared to a classifier trained on a clean-labeled dataset. This study highlights the importance of refining noisy-labeled data for training MSS models effectively and demonstrates that utilizing the refined dataset for MSS yields results comparable to those obtained using a clean-labeled dataset. Considering the real-world scenario of accessibility only to a noisy dataset, MSS models trained on self-refined datasets outperformed those trained on datasets refined with a classifier trained on clean labels. The self-refining approach we introduced offers a promising direction for future research in the field of music information retrieval and has the potential to be extended to other applications requiring robust training on noisy-labeled datasets.

\section{Acknowledgements}
\label{section:ack}
This work was partially supported by Culture, Sports and Tourism R\&D Program through the Korea Creative Content Agency grant funded by the Ministry of Culture, Sports and Tourism in 2022 [No.R2022020066, 90\%], and Institute of Information \& communications Technology Planning \& Evaluation (IITP) grant funded by the Korea government(MSIT) [NO.2021-0-01343, Artificial Intelligence Graduate School Program (Seoul National University), 10\%].

\bibliography{ISMIRtemplate}

\onecolumn
\newpage
\renewcommand\thesection{\Alph{section}}
\setcounter{section}{0}

\section{Iterative Self-refining}
We further investigate the impact of the iterative self-refining technique on both multi-instrument recognition and MSS performance. From Table \ref{table:inst_cla_obj_results_iterative_refining}, we observe that the multi-instrument recognition performance tends to increase with additional self-refining iterations, although not significantly. In the context of music source separation, we perform the same configuration as in Section \ref{subsubsection:ablation}, where we train and evaluate the impact upon iterative self-refinement with the X-UMX. Interestingly, we do not observe an improvement in MSS performance as the self-refining iterations increase from Table \ref{table:mss_results_iterative_refining}.
Considering training efficiency, we find that self-refining the dataset twice is sufficient for optimal performance with the MDX2023 dataset, though this may vary with other noisy-labeled datasets. Additionally, since the percentage of corruption in the MDX 2023 dataset is unknown, further analysis on the correlation between the percentage of noisy labels and the effectiveness of self-refining remains challenging.

\begin{table}[h]
\centering
\begin{tabular}{ccccccc}
\hline
\multirow{2}{*}{\textbf{Label Type}} & \multirow{2}{*}{\textbf{\begin{tabular}[c]{@{}c@{}}Self-refining \\ Iteration\end{tabular}}} & \multicolumn{5}{c}{\textbf{\begin{tabular}[c]{@{}c@{}}Accuracy / F1 Score\\ Precision / Recall\end{tabular}}}                                                                                                                                                                                                                                                                                                   \\ \cline{3-7} 
                                     &                                         & \textit{vocals}                                                               & \textit{bass}                                                                 & \textit{drums}                                                                & \textit{other}                                                                  & \textit{avg}                                                                  \\ \hline
\multirow{11}{*}{Single-Label}        & \textit{clean}                          & \begin{tabular}[c]{@{}c@{}}97.8\% / 0.947\\ 0.91 / 0.98\end{tabular}          & \begin{tabular}[c]{@{}c@{}}94.4\% / 0.891\\ 0.84 / 0.94\end{tabular}          & \begin{tabular}[c]{@{}c@{}}95.1\% / 0.914\\ 0.85 / 0.98\end{tabular}          & \begin{tabular}[c]{@{}c@{}}93.2\% / 0.880\\ 0.90 / 0.85\end{tabular}            & \begin{tabular}[c]{@{}c@{}}95.1\% / 0.906\\ 0.87 / 0.93\end{tabular}          \\ \cdashline{2-7}
                                     & \textit{$\times1$}                          & \begin{tabular}[c]{@{}c@{}}93.6\% / 0.860\\ 0.76 / 0.97\end{tabular}          & \begin{tabular}[c]{@{}c@{}}90.0\% / 0.821\\ \textbf{0.73} / 0.93\end{tabular} & \begin{tabular}[c]{@{}c@{}}\textbf{93.7\%} / \textbf{0.893}\\ \textbf{0.81} / 0.98\end{tabular} & \begin{tabular}[c]{@{}c@{}}92.6\% / 0.865\\ \textbf{0.92} / 0.81\end{tabular} & \begin{tabular}[c]{@{}c@{}}92.5\% / 0.860\\ 0.80 / 0.92\end{tabular}          \\
                                     & \textit{$\times2$ (proposed)}                        & \begin{tabular}[c]{@{}c@{}}96.1\% / 0.911\\ 0.84 / \textbf{0.98}\end{tabular} & \begin{tabular}[c]{@{}c@{}}89.6\% / 0.818\\ 0.71 / \textbf{0.96}\end{tabular}          & \begin{tabular}[c]{@{}c@{}}93.1\% / 0.884\\ 0.79 / 0.98\end{tabular}          & \begin{tabular}[c]{@{}c@{}}92.3\% / 0.862\\ 0.90 / 0.82\end{tabular}            & \begin{tabular}[c]{@{}c@{}}92.8\% / 0.866\\ 0.80 / \textbf{0.93}\end{tabular} \\
                                     & \textit{$\times3$}                        & \begin{tabular}[c]{@{}c@{}}95.8\% / 0.904\\ 0.83 / \textbf{0.98}\end{tabular} & \begin{tabular}[c]{@{}c@{}}90.5\% / 0.830\\ \textbf{0.73} / 0.95\end{tabular}          & \begin{tabular}[c]{@{}c@{}}93.2\% / 0.886\\ 0.80 / 0.98\end{tabular}          & \begin{tabular}[c]{@{}c@{}}91.9\% / 0.851\\ 0.91 / 0.79\end{tabular}            & \begin{tabular}[c]{@{}c@{}}92.8\% / 0.866\\ \textbf{0.81} / 0.92\end{tabular} \\
                                     & \textit{$\times4$}                        & \begin{tabular}[c]{@{}c@{}}95.1\% / 0.890\\ 0.81 / \textbf{0.98}\end{tabular} & \begin{tabular}[c]{@{}c@{}}\textbf{90.7\%} / \textbf{0.836}\\ \textbf{0.73} / \textbf{0.96}\end{tabular}          & \begin{tabular}[c]{@{}c@{}}92.6\% / 0.877\\ 0.78 / \textbf{0.99}\end{tabular}          & \begin{tabular}[c]{@{}c@{}}92.3\% / 0.859\\ 0.91 / 0.81\end{tabular}            & \begin{tabular}[c]{@{}c@{}}92.7\% / 0.864\\ 0.80 / 0.93\end{tabular} \\
                                     & \textit{$\times5$}                        & \begin{tabular}[c]{@{}c@{}}\textbf{96.9\%} / \textbf{0.927}\\ \textbf{0.88} / 0.97\end{tabular} & \begin{tabular}[c]{@{}c@{}}90.4\% / 0.830\\ \textbf{0.73} / 0.95\end{tabular}          & \begin{tabular}[c]{@{}c@{}}92.3\% / 0.873\\ 0.78 / \textbf{0.99}\end{tabular}          & \begin{tabular}[c]{@{}c@{}}\textbf{92.8\%} / \textbf{0.872}\\ 0.90 / \textbf{0.83}\end{tabular}            & \begin{tabular}[c]{@{}c@{}}\textbf{93.1\%} / \textbf{0.872}\\ \textbf{0.81} / \textbf{0.93}\end{tabular} \\ \hline

\multirow{11}{*}{Multi-Label}         & \textit{clean}                          & \begin{tabular}[c]{@{}c@{}}92.4\% / 0.929\\ 0.92 / 0.93\end{tabular}          & \begin{tabular}[c]{@{}c@{}}89.6\% / 0.905\\ 0.89 / 0.92\end{tabular}          & \begin{tabular}[c]{@{}c@{}}90.5\% / 0.913\\ 0.87 / 0.95\end{tabular}          & \begin{tabular}[c]{@{}c@{}}88.1\% / 0.878\\ 0.90 / 0.85\end{tabular}            & \begin{tabular}[c]{@{}c@{}}90.2\% / 0.907\\ 0.90 / 0.91\end{tabular}          \\ \cdashline{2-7}
                                     & \textit{$\times1$}                          & \begin{tabular}[c]{@{}c@{}}87.9\% / 0.895\\ 0.83 / 0.96\end{tabular}          & \begin{tabular}[c]{@{}c@{}}87.5\% / 0.888\\ \textbf{0.86} / 0.93\end{tabular}          & \begin{tabular}[c]{@{}c@{}}87.7\% / 0.891\\ 0.82 / 0.96\end{tabular}          & \begin{tabular}[c]{@{}c@{}}87.3\% / 0.872\\ 0.88 / \textbf{0.87}\end{tabular}            & \begin{tabular}[c]{@{}c@{}}87.6\% / 0.887\\ 0.85 / 0.93\end{tabular}          \\
                                     & \textit{$\times2$ (proposed)}                        & \begin{tabular}[c]{@{}c@{}}91.9\% / 0.928\\ 0.88 / \textbf{0.97}\end{tabular} & \begin{tabular}[c]{@{}c@{}}87.8\% / 0.894 \\ 0.84 / \textbf{0.95}\end{tabular} & \begin{tabular}[c]{@{}c@{}}\textbf{89.6\%} / \textbf{0.906}\\ \textbf{0.85} / 0.96\end{tabular} & \begin{tabular}[c]{@{}c@{}}87.4\% / 0.874\\ 0.88 / \textbf{0.87}\end{tabular}   & \begin{tabular}[c]{@{}c@{}}89.2\% / 0.901\\ 0.86 / \textbf{0.94}\end{tabular} \\
                                     & \textit{$\times3$}                        & \begin{tabular}[c]{@{}c@{}}91.5\% / 0.924\\ 0.88 / 0.96\end{tabular} & \begin{tabular}[c]{@{}c@{}}89.1\% / 0.903\\ \textbf{0.86} / 0.94\end{tabular}          & \begin{tabular}[c]{@{}c@{}}89.3\% / 0.903\\ \textbf{0.85} / 0.96\end{tabular}          & \begin{tabular}[c]{@{}c@{}}87.9\% / 0.874\\ \textbf{0.90} / 0.84\end{tabular}            & \begin{tabular}[c]{@{}c@{}}89.5\% / 0.902\\ \textbf{0.87} / 0.93\end{tabular} \\
                                     & \textit{$\times4$}                        & \begin{tabular}[c]{@{}c@{}}90.7\% / 0.917\\ 0.87 / 0.96\end{tabular} & \begin{tabular}[c]{@{}c@{}}\textbf{89.4\%} / \textbf{0.906}\\ \textbf{0.86} / \textbf{0.95}\end{tabular}          & \begin{tabular}[c]{@{}c@{}}88.6\% / 0.898\\ 0.83 / \textbf{0.97}\end{tabular}          & \begin{tabular}[c]{@{}c@{}}87.6\% / 0.871\\ \textbf{0.90} / 0.84\end{tabular}            & \begin{tabular}[c]{@{}c@{}}89.1\% / 0.899\\ 0.86 / 0.93\end{tabular} \\
                                     & \textit{$\times5$}                        & \begin{tabular}[c]{@{}c@{}}\textbf{92.5\%} / \textbf{0.932}\\ \textbf{0.90} / 0.96\end{tabular} & \begin{tabular}[c]{@{}c@{}}88.8\% / 0.901\\ 0.85 / 0.94\end{tabular}          & \begin{tabular}[c]{@{}c@{}}89.0\% / 0.901\\ 0.84 / \textbf{0.97}\end{tabular}          & \begin{tabular}[c]{@{}c@{}}\textbf{88.1\%} / \textbf{0.877}\\ \textbf{0.90} / 0.85\end{tabular}            & \begin{tabular}[c]{@{}c@{}}\textbf{89.6\%} / \textbf{0.904}\\ \textbf{0.87} / 0.93\end{tabular} \\ \hline
\end{tabular}
\caption{
    Single and multi-label instrument recognition performance of instrument classifiers self-refined with the \textit{noisy} (MDX2023) dataset for $n$ times.
}
\label{table:inst_cla_obj_results_iterative_refining}
\end{table}

\begin{table}[h]
\centering
\begin{tabular}{lccccc}
\hline
\multirow{3}{*}{\textbf{Training Data}} & \multicolumn{5}{c}{\multirow{2}{*}{\textbf{SDR {[}dB{]}}}} \\
 & \multicolumn{5}{l}{} \\ \cline{2-6} 
 & \textit{vocals} & \textit{bass} & \textit{drums} & \textit{other} & \textit{avg} \\ \hline
\textit{clean} & 5.76 & 4.44 & 5.47 & 3.65 & 4.83 \\ \cdashline{1-6}
\textit{noisy ($\times0) $} & 3.39 & 1.78 & 1.52 & 0.96 & 1.91 \\ \cdashline{1-6}
\textit{$\times1$} & 4.72 & \textbf{4.11} & 3.22 & 2.89 & 3.74 \\
\textit{$\times2$ (proposed)} & 4.99 & 3.93 & 5.00 & \textbf{3.18} & \textbf{4.28} \\
\textit{$\times3$} & \textbf{5.03} & 3.81 & 4.05 & 3.02 & 3.98 \\
\textit{$\times4$} & 4.86 & 3.81 & 4.64 & 3.07 & 4.09 \\
\textit{$\times5$} & 4.65 & 3.87 & \textbf{5.07} & 2.89 & 4.12 \\ \hline
\end{tabular}
\caption{
    MSS performances of CrossNet-Open-Unmix trained with the datasets self-refined for $n$ times. 
    The notation $\times n$ indicates the \textit{noisy} (MDX2023) dataset being inferred with the classifier trained using $\times (n-1)$.
}
\label{table:mss_results_iterative_refining}
\end{table}

\end{document}